\documentclass[a4paper, amsfonts, amssymb, amsmath, pra, reprint, showkeys, nofootinbib, 
]{revtex4-1}
\usepackage[english]{babel}
\usepackage[utf8]{inputenc}
\usepackage[colorinlistoftodos, color=green!40, prependcaption]{todonotes}
\usepackage[pdftex, pdftitle={Article}, pdfauthor={Author}]{hyperref} 
\usepackage{soul}
\usepackage{graphicx}
\usepackage[T1]{fontenc} 
\usepackage{bm,latexsym,amsmath,amssymb,amsfonts}
\usepackage{xcolor}
\usepackage{ulem}

\newcommand{\be}[1]{\begin{equation} \label{#1}}
\newcommand{\ee}{\end{equation}}
\newcommand{\bea}{\begin{eqnarray}}
\newcommand{\eea}{\end{eqnarray}}
\newcommand{\nn}{\nonumber}
\newcommand{\hreff}[1]{\href{#1}{\color{blue}{#1}} }

\begin{document}

\title{Electromagnetism from relativistic fluid dynamics }

\author{Jeongwon Ho}
        \email{freejwho@gmail.com}
        \affiliation{Department of Physics, Institute of Basic Science, Sungkyunkwan University, Suwon 16419, Korea}

\author{Hyeong-Chan Kim}
        \email{Corresponding Author, hckim@ut.ac.kr}
	\affiliation{School of Liberal Arts and Sciences, Korea National University of Transportation, Chungju 27469, Korea}

\author{Jungjai Lee}
	\email{jjlee@daejin.ac.kr}
	\affiliation{Graduate School Department of Physics, Daejin University, Pocheon 11159, Korea}
	\affiliation{School of Physics, Korea Institute for Advanced Study, Seoul 02455, Korea}
\author{Yongjun Yun}
	\email{yongjun5828@gmail.com}
	\affiliation{Graduate School Department of Physics, Daejin University, Pocheon 11159, Korea}

\date{\today} 

\begin{abstract}
We reformulate classical electromagnetism within the matter-space framework of relativistic fluid dynamics. 
The central assumption is that the relevant degrees of freedom are encoded in differential forms on a three-dimensional matter space and mapped to spacetime by pull-back.
The absence of four-forms in matter space imposes nontrivial kinematical constraints on the induced spacetime fields and restricts gauge transformations to those compatible with the flow. 
Because of this (matter-space) gauge symmetry, the physically relevant sector is retained, and the Aharonov-Bohm phase is naturally associated with the matter-space potential. 
The construction admits two electromagnetic frames. 
We argue that the frame identifying the spacetime field strength directly with the intrinsic matter-space two-form is geometrically preferred. 
In the first frame, the homogeneous sector is fixed by the matter-space structure, while the sourced equation follows from an action-based
relativistic-fluid formulation in a first-order setting where the potential and field strength are varied independently and the matter-space constraints are imposed on shell. 
In the massless case and to quadratic order, locality and the (matter-space) gauge symmetry fix the leading field term in the action uniquely, so the resulting equations provide the minimal dynamical completion once charge carriers are included.
We also clarify how duality controls the status of the Bianchi
identity in the absence of magnetic charge carriers, and we briefly discuss helicity conservation and a natural nonlinear extension implied by the one-fluid constraints.
In the second frame, on the other hand, the matter space 1-form is not directly related with the gauge potential.
\end{abstract}

\keywords{electromagnetism, relativistic fluid dynamics, matter-space formulation}

\maketitle
\onecolumngrid
 
\section{Introduction}
Astronomical magnetic fields play crucial roles in the formation and evolution of galaxies, stars, and planets. 
Their origin and amplification are generally explained by the dynamo mechanism (for a comprehensive review, see \cite{Brandenburg}), providing a theoretical framework for understanding the galactic magnetic fields observed today \cite{Subramanian}. 
Magnetohydrodynamics (MHD) uses the electromagnetism and fluid dynamics to describe the interaction between electromagnetic fields and conductive fluids \cite{Davidson}. 
Despite its success in providing phenomenological explanations, MHD faces significant challenges. 
Observational evidence for the initial seed magnetic fields remains elusive, and the theoretical mechanisms proposed for their generation are not yet fully understood \cite{Kulsrud}. 
Furthermore, critical issues persist, such as understanding how the back-reaction of magnetic fields limits magnetic amplification in turbulent flow and the mechanisms reasonable for the generation/maintenance of large-scale galactic magnetic fields \cite{Pariev}. 
These challenges partly stem from the fact that, MHD combines hydrodynamic effective/collective/coarse-grained descriptions of matter with a fundamental gauge-field description of electromagnetism arising from microscopic physics of constituent charged particles.
 As a result, different theoretical frameworks are employed, leading to a lack of a unified approach, which poses a major obstacle to achieving a consistent and comprehensive interpretation of the underlying physics.
 
In this paper, we propose a novel theoretical approach to address these challenges. Our research aims to develop a unified framework that integrates electromagnetism and fluid dynamics, thereby transcending their conventional separate treatment in MHD. We anticipate that this unified framework will not only provide a more fundamental understanding of the interaction between electromagnetic fields and fluids but also offer a relativistic extension necessary for explaining high-energy astrophysical phenomena \cite{Hernandez} such as supernova explosions, gamma-ray bursts, and black hole jets in addition to nuclear fusion.

Before delving further, it is important to outline a theoretical motivation for this study. 
In modern physics, particles can be interpreted as excitations of quantum fields  {\it inhabiting spacetime}, endowed with specific symmetries and properties such as spin, charge, and mass.
 The field-theoretic approach to matter and particles originates from electromagnetism, where electric and magnetic interactions were unified within the framework of a local $U(1)$ gauge theory. Quantum field theory, built on this foundation, has been remarkably successful in describing fundamental interactions, including electromagnetic, weak, and strong forces. This framework has also been applied successfully to gravity, at least in classical regime. However, limitations in existing paradigms, such as the treatment of singularities, have driven the exploration of alternative frameworks like string theory \cite{Green}, loop quantum gravity \cite{Rovelli}, and the emergent point of view for gravity \cite{Jacobson,Padmanabhan,Lee}.

On the other hand, the relativistic fluid and thermodynamic approach assumes that matter or particles reside in a three-dimensional Euclidean space, referred to as ``matter space,'' with their dynamics governed by mappings between this matter space and four-dimensional spacetime, illustrated in Fig. 1,  \cite{Taub54,Carter72,Carter73,Carter89,AnderssonNew,LK2022,Kim:2023lta,Dubovsky:2011sj}. 

\begin{figure}[h]
    \centering
    \includegraphics[width=1\linewidth]{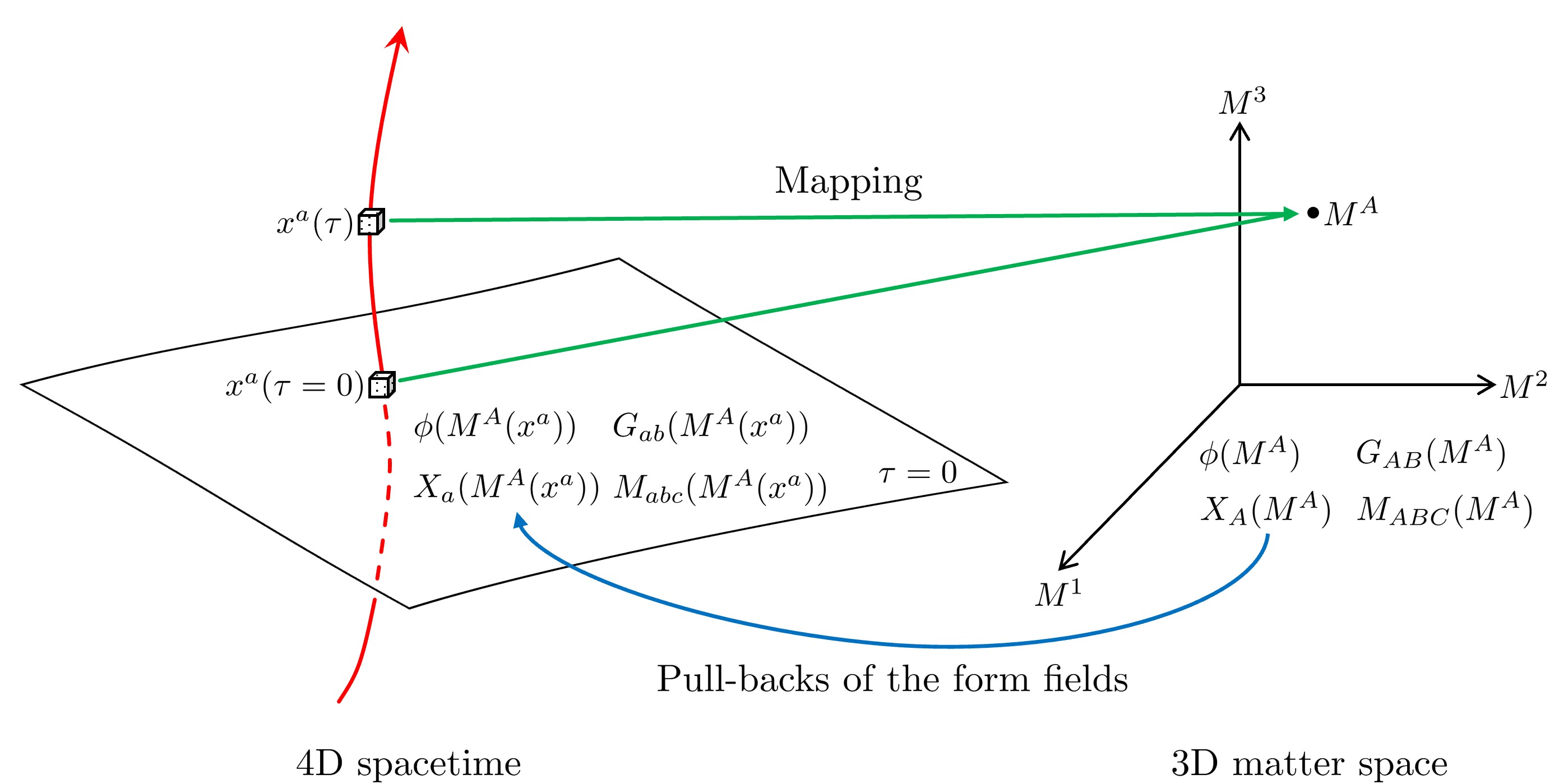}
    \caption{An illustration of the pull-back description from fluid-particle points in the three-dimensional matter space, labelled by the coordinates $\{M^1, M^2, M^3\}$, to fluid-element worldlines in spacetime. 
The pull-back of a fluid particle to, say, an initial point on a worldline in spacetime can be taken as  $M^A = M^A(0, x^i)$ where $(0,x^i)$ is the spacetime point of the intersection of the worldline with the $t=0$ time slice.
Here $\{\phi, X_A, G_{AB},M_{ABC} \}$ and $\{\phi, X_a, G_{ab}, M_{abc}\}$ are the 0, 1, 2, 3-form fields residing on the matter space and spacetime, respectively. 
 }
    \label{fig:sample}
\end{figure}

In this work, we adopt this perspective by considering matter (specifically, photons) to reside in a three-dimensional matter space, while the electromagnetic fields observed in four-dimensional spacetime represent projections of this underlying system.
Our approach naturally accommodates configurations with magnetic-source-like structure while preserving electromagnetic duality. Consequently, the Aharonov--Bohm effect is associated with the matter-space gauge potential, while helicity conservation emerges in a natural way. Moreover, nonlinear corrections analogous to mass renormalization are intrinsically incorporated. From a broader conceptual standpoint, our approach establishes a unified foundation for treating electromagnetism and fluid dynamics within a single framework, thereby offering new insight into the interplay between matter and electromagnetic fields. Our framework suggests that the dynamics of electromagnetic fields can be understood as a pull-back mechanism, illustrated in Fig.~\ref{fig:sample}, from matter space to spacetime.

In Sec.~\ref{sec2}, we introduce the matter-space construction and the associated constraints on form fields pulled back from a three-dimensional matter space to four-dimensional spacetime. We then discuss the two admissible duality frames and identify the frame $F_{ab}=G_{ab}$ as the geometrically preferred one for the main development of the theory. In Sec.~\ref{sec3}, we derive the remaining Maxwell equation through the standard action-based formulation of relativistic fluid dynamics, treating the constituent form fields as independent variables. Upon imposing the matter-space constraints, this variational formulation yields the electric-source sector and recovers the standard sourced Maxwell equation. In Sec.~\ref{sec4}, we analyze the role of electromagnetic duality in clarifying the status of the Bianchi identity and the branch with magnetic-type sources, especially in the absence of external magnetic charge carriers. In Sec.~\ref{sec5}, we examine the alternative  frame $F_{ab}=-\mbox{*}G_{ab}$ for comparison. Finally, in Sec.~\ref{sec6}, we summarize our principal results and discuss several conceptual and physical implications of the framework.

\section{Gauge symmetry and matter space formulation} \label{sec2}

In this work, we consider a one-fluid system incorporating a single matter space flowing along $m^a$ defined below. 
Let the three-dimensional matter space have coordinates \(-\infty < M^A < \infty\), where \(A = 1, 2, 3\), as illustrated in Fig.~\ref{fig:sample}. 
We emphasize that {\it no metric structure} is assumed on the matter space. In particular, there is no notion of Hodge duality intrinsic to the matter space, and differential forms of different degrees must therefore be treated as independent objects.
The coordinates \(M^A(x^a)\) co-move with their respective world-lines in four-dimensional spacetime.
We are interested only in form fields to be associated with matter, which are anti-symmetric tensor fields, rather than symmetric ones. 
Since the matter space is three-dimensional, the following form fields may exist: a 0-form field \(\phi(M^A)\), a 1-form field \(\bm{X} = X_A (M^A)\, dM^A\), a 2-form field \(\bm{G} = \frac{1}{2!} G_{AB}(M^A) \, dM^A \wedge dM^B\), and a 3-form field \(\bm{M}=\frac{1}{3!} M_{ABC}(M^A) \, dM^A \wedge dM^B \wedge dM^C\). 
Each form encodes distinct physical information associated with the matter degrees of freedom.
The fundamental conjecture in this work is that these forms characterize the matter itself.

In fluid dynamics the 3-form field $\bm{M}$ typically encodes the direction of fluid flow. 
Its spacetime dual, defined as   
\be{ma}
m^a \equiv \frac{1}{3!} \epsilon^{abcd}M_{bcd},
\ee
corresponds to the flow vector of the fluid, where the 3 form $M_{abc}$ is the spacetime projection of the matter space field, 
$M_{abc} =(\nabla_a M^A)(\nabla_b M^B) (\nabla_c M^C)M_{ABC}$.
The appearance of the dual vector $m^a$
 relies on the {\it spacetime metric} and Levi-Civita tensor, and should be regarded as an emergent spacetime construct rather than a fundamental matter-space field.
In this sense, when we construct electromagnetism based on the dynamics of a single matter space, the results must inherently possess a directional structure aligned with $m^a$.
In ordinary fluid dynamics, the flow field $m^a$ is often interpreted as encoding the conservation of particle number.  
However, for the case of the electromagnetic field, the number density may not be necessarily conserved because the photon is massless.
In this sense, the conservation law 
\be{conservm}
\nabla_a m^a=0
,
\ee
which comes from $\bm{d M}=0$, does not imply particle number conservation.
Since we are interested in electromagnetism, we begin with the 1- and 2-form fields and later identify $m^a$ as a kind of vector denoting helicity flux or electric current.

Guided by this perspective, we reconstruct electromagnetism from the behavior of the matter-space form fields. The corresponding spacetime fields, induced through the geometric mapping depicted in Fig.~\ref{fig:sample}, are 
\begin{align}    \label{inducedstfields}
\phi(x^a) &\equiv \phi(M^A(x^a)),  \nonumber \\
X_a(x^a) & \equiv (\nabla_a M^A(x^a)) X_A(M^B(x^a)),    \\
G_{ab}(x^a) &\equiv (\nabla_a M^A(x^a))(\nabla_b M^B(x^a)) G_{AB}(M^C(x^a)), \nonumber
\end{align}
in addition to the 3-form $M_{abc}$ introduced in Eq.~\eqref{ma}. It is important to note that $X_a$ is treated here as an independent 1-form field and is not a priori assumed to be the conventional electromagnetic gauge potential.

Before proceeding further, we clarify how the matter-space gauge symmetry is realized in this framework.
Since our interest lies in the matter space flowing along $m^a$, the induced fields obey specific constraints. 
Because the matter space has lower dimensionality than the physical spacetime, relations that are trivial within the matter space can yield nontrivial consequences once mapped to spacetime. 
For instance, in the three-dimensional matter space, the wedge product $\bm{M} \wedge \bm{X}$ vanishes identically, as no 4-forms exist.
However, upon projection to spacetime, one finds 
\begin{align} \label{M^A2}
0= \bm{M} \wedge \bm{X} 
= & \frac{1}{3!} M_{ABC} X_{D} dM^A \wedge dM^B \wedge dM^C \wedge dM^D \nonumber \\
=& \frac{1}{3!} M_{abc} X_{d} dx^a \wedge dx^b \wedge dx^c \wedge dx^d \nn \\
=& \frac{1}{3!} M_{abc} X_{d} \epsilon^{abcd} \times (\textbf{Vol})  = -m^d X_d  \times (\textbf{Vol}) 
\end{align}
where $(\textbf{Vol})$ denotes the 4-dimensional volume form.
Here, Eq.~\eqref{M^A2} implies the non-trivial condition,
\be{M^A}
m^a X_a =0.
\ee
Thus, the 1-form $X_a$ is restricted to directions orthogonal to $m^a$.
The corresponding field strength $\bm{d X}$ must remain invariant under a transformation $X_a\to X'_a = X_a + \nabla_a \phi'$. 
However, if the scalar field $\phi'$  satisfies $m^a \nabla_a \phi' \neq 0$, the transformed 1-form fails to remain within the matter-space form, as then $m^a X'_a \neq 0$.  
Accordingly, scalar fields naturally split into two classes:
i) $\phi'$, which displace the potential outside the matter space; 
ii) $\phi$, which preserve the restriction and thus generate the matter-space gauge symmetry, characterized by 
\be{m^dphi} 
m^a \nabla_a \phi=0.
\ee
This condition arises equivalently from the matter-space relation $\bm{M} \wedge \bm{d \phi}=0$.
Hence, the matter-space scalar $\phi$ encodes the gauge degrees of freedom, which are pulled back from the matter space. 
By contrast, scalars with $m^a \nabla_a \phi' \neq 0$ belong to the first class and do not generate admissible transformations.

Therefore, within the matter space formulation, the full gauge symmetry of electromagnetism does not survive; rather, it is restricted to the matter-space symmetry obeying~\eqref{m^dphi}.
This restriction reflects the fact that the potential is constructed from matter-space structures transported along $m^a$.
As we show later, despite this restriction, it remains possible to  construct electromagnetism consistently by invoking the duality between electric and magnetic fields. Finally, from the definition of $m^a$, we find 
\begin{equation} \label{orthom}
 m^a G_{ab} = 0,
\end{equation}
restricting the gauge field to be normal to $m^a$ as well.

At this point, we need to examine the degrees of freedom of the matter space. 
Fundamentally, the matter space has three degrees of freedom consisting of $M^A$ with $A=1,2,3$. 
On the other hand, in four-dimensional spacetime, the electromagnetic potential $A_a$ and fields $F_{ab}$ consist of four and six variables, respectively. 
The gauge potential is restricted to satisfy \eqref{M^A}, leaving three degrees of freedom; this restriction changes not the physics but the gauge description, since observables are represented by the field strength.
The 2-form field, given the direction $m^a$, also has three freedoms satisfying $m^a G_{ab}=0$\footnote{ 
As a specific example, one can choose a coordinate system in which the time coordinate follows the integral curves of $m^a$, so that $m^a = m^0 \delta^a_0$. In this frame, the condition in Eq. \eqref{orthom} implies 
$
G_{0k} =0 .
$
Now, the form field has three non-vanishing degrees of freedom, $G_{12}$, $G_{23}$, $G_{31}$.}.
In this sense, the 6 freedoms of the electromagnetic field is redistributed as 3+3 for the direction of $m^a$ and for the matter space 2-form field, respectively.
If we are to construct a general electromagnetic field based on the matter-space approach, we need to consider multi-fluid theory with many propagating directions, $m^a $ vectors.

As previously mentioned, we interpret $\phi$ as the scalar field encoding the gauge degrees of freedom.
Let us now examine its implications for the 2-form field $\bm{G}$.
The vanishing of the 4-form,
\begin{align}
0 &= \bm{d\phi} \wedge \bm{dG}= \frac{1}{4!} (\nabla_{[a} \phi) (\nabla_b G_{cd]}) dx^a \wedge dx^b \wedge dx^c \wedge dx^d,
\label{dphi:dF}
\end{align}
yields the nontrivial relation:
\begin{align}\label{DphiDF}
\epsilon^{abcd} (\nabla_{[a} \phi)(\nabla_b G_{cd]}) = 0. 
\end{align}
Because $\phi$ represents the gauge degrees of freedom, this identity must hold for an arbitrary scalar field $\phi$ satisfying Eq.~\eqref{m^dphi}. 
Expressing Eq.~\eqref{DphiDF} in terms of the dual tensor $\mbox{*}G^{ab} \equiv \frac12 \epsilon^{abcd}G_{cd}$, we obtain
\bea \label{DphiDF1} 
\epsilon^{abcd} (\nabla_{[a} \phi)(\nabla_b G_{cd]}) 
	&=&(\nabla_a\phi) (\nabla_b \,*G^{ab}) =  0
 \\
&&
 \forall \phi \mbox{ satisfying } m^a \nabla_a \phi =0 .  \nn 
\eea 
Given the constraint $m^a \nabla_a \phi =0$, this requirement implies that    
\be{*dG}
\nabla_{b} \mbox{*}G^{ab} = \alpha m^a ,
\ee
where $\alpha$ is a scalar function\footnote{In general, $\alpha$ may be position-dependent. However, for the theory to be consistently formulated within the matter-space formalism, this dependence must be described by a gauge-independent scalar quantity constructed entirely from matter-space fields.}.
We emphasize that this is the only sourced equation arising from the kinematical constraints of the matter space. 

The electromagnetic field strength $F_{ab}$ may be identified either with the Hodge dual $-\mbox{*}G_{ab}$ or directly with $G_{ab}$, corresponding to two distinct frames:
\be{F:G}
 F_{ab}  = G_{ab}\quad \mbox{or} \quad F_{ab} = -\mbox{*}G_{ab} . 
\ee
These two identifications are related by electromagnetic duality, but they assign different physical roles to the sourced equation~\eqref{*dG}:
\be{*dF}
\nabla_{b}\mbox{*}F^{ab} = \alpha m^a
\quad \mbox{or} \quad 
\nabla_b F^{ab} = \alpha m^a   .
\ee
Accordingly, the term $\alpha m^a$ is interpreted as a magnetic source in the first frame and as an electric source in the second. 

Although both frames are formally admissible, the identification $F_{ab}=G_{ab}$ is more directly compatible with the matter-space construction, because it keeps the physical field strength tied to the intrinsic matter-space 2-form itself rather than to its spacetime Hodge dual. 
For this reason, we first develop the theory in the frame $F_{ab}=G_{ab}$ in this section and return later to the alternative identification $F_{ab}=-\mbox{*}G_{ab}$ for comparison.

\vspace{.3cm}
We first regard the electromagnetic field $F_{ab}$ as the pull-back of the closed 1-forms $G_{AB}$ in matter space satisfying $\bm{dG}=0$.
Then, the standard homogeneous Maxwell equation, i.e., the Bianchi identity,
\begin{equation} \label{Maxeq1}
 \nabla_{[a}F_{bc]} = 0  ~.
\end{equation}
appears naturally, which forces $\alpha =0$ in Eq.~\eqref{*dF}.

On the other hand, when $F_{ab}$ is a pull-back of general matter space 2-forms which may not be closed, the Bianchi identity may not hold. 
In that case,
Eq.~\eqref{*dF} becomes an inhomogeneous equation for the dual field $\mbox{*}\bm{F}$, so that $\alpha m^a$ is naturally interpreted as a magnetic source current $j_M^a \equiv \alpha m^a$\footnote{
Let us reconsider the previous example involving a constant timelike vector $m^a$.
The conditions from Eqs.~\eqref{M^A}, \eqref{m^dphi}, and \eqref{orthom} imply that $\phi$ must be time-independent, $A_0 = 0$, and $F_{0k} = 0$.
Nonetheless, $\phi$ may depend arbitrarily on the spatial coordinates. From Eq.~\eqref{DphiDF}, one deduces $\nabla_{[0} F_{jk]} = 0$, indicating that the magnetic field is static. The condition $\epsilon^{abcd} A_a (\nabla_{[b} F_{cd]}) = 0$ is also satisfied under these assumptions. Furthermore, from Eq.~\eqref{*dG}, one finds in this frame:
$$
\nabla_{[i} F_{jk]} = -\frac{2}{3} \alpha  \epsilon_{ijk0} m^0 .
$$ 
Eq.~\eqref{*dG} also yields $\nabla_{[0} F_{jk]} = 0$. 
Physically, because $\phi$ is constant in time and $F_{0k} = 0$, the electric field vanishes, leaving only a static magnetic configuration. The relation $\nabla_{[i} F_{jk]} \neq 0$ directly implies $\nabla\cdot \vec{B} \neq 0$, demonstrating that $m^a$ effectively plays the role of a magnetic source.}.
The implications of allowing $\bm{dG}\neq 0$, and the extent to which this possibility is compatible with electromagnetic duality, will be examined more fully in Sec.~\ref{sec4}.

In domains where the source vanishes, the closed 2-form $F_{bc}$ can be written locally as \(F_{bc} = \nabla_b A_c - \nabla_c A_b\), parameterized by a 1-form field \(\bm{A}\).  
In a similar vein, the matter-space relations $\bm{d X} \wedge \bm{G} = 0$ and $\bm{X} \wedge \bm{dG} = 0$ yield the following spacetime constraints:
\begin{eqnarray} \label{orthoDAF}
\epsilon^{abcd} (\nabla_{[a} X_{b]} ) F_{cd}  = \mbox{*}F^{ab} (\nabla_{[a} X_{b]} ) = 0 , \nonumber \\
\epsilon^{abcd}  X_{a} \nabla_{[b} F_{cd]} = 0 .
\end{eqnarray}
In other words, $ \mbox{*}F_{ab}$ is orthogonal to $ \nabla_{[a} X_{b]}$, i.e., $\mbox{*}F^{ab} (\nabla_{[a} X_{b]} ) =0 $. Furthermore, the constraint equation, $\bm{G} \wedge \bm{G} = 0$, which is also a 4-form field, presents
\begin{align} \label{tildeff}
\mbox{*}F^{ab}F_{ab} =   \epsilon^{abcd} F_{ab} F_{cd} = 0.
\end{align}
This equation implies $\bm{F\wedge F}=0$, i.e. $\vec{E}\cdot \vec{B}=0$ in a $3+1$ split.  Hence the minimal one-fluid construction selects the degenerate sector of electromagnetism; relaxing this restriction generally requires multiple matter-space flows.
To summarize, the spacetime-induced fields satisfy the relations in Table~\ref{table1}\footnote{In deriving the equations listed in this table, we utilized the fact that all 4-forms identically vanish within the 3-dimensional matter space. Note that the vanishing of 5-forms or higher does not yield new constraints, as they are identically zero in both the matter space and the 4-dimensional spacetime.}.

\begin{table}[h]
\centering
\begin{tabular}{@{} c c @{}}
\hline  
\text{(A)} & \text{(B)} \\
\hline  
$\epsilon^{abcd} (\nabla_{[a} \phi)(\nabla_b F_{cd]}) = 0$ & \quad
$\epsilon^{abcd} (\nabla_{[a} X_{b]}) F_{cd} = 0$ \\
\hline  
\text{(C)} & \text{(D)} \\
\hline  
$\epsilon^{abcd} X_a (\nabla_{[b} F_{cd]}) = 0$ & \quad
$\epsilon^{abcd} F_{ab} F_{cd} = 0$ \\
\hline  
\end{tabular}
\caption{Relation for spacetime-induced fields}
\label{table1}
\end{table}

Whenever the Bianchi identity~\eqref{Maxeq1} holds, relation (C) in Table~\ref{table1} is satisfied automatically. The remaining nontrivial information is therefore carried by relations (B) and (D), which impose strong constraints linking the two 1-forms $X_a$ and $A_a$.

In general, the 1-form field $\bm{A}$, which plays the role of the usual gauge potential, need not automatically reside within the matter space. Explicitly, even if $A_a$ is initially a matter-space form, adding a constant vector with a component along the $m^a$ direction would violate this restriction. However, one can always perform a gauge transformation, choosing a scalar $\phi'$ such that the transformed potential $A_a' = A_a + \nabla_a \phi'$ satisfies $m^a A_a' = 0$.

Since our foundational premise is that the essential properties of matter must be described entirely by matter-space fields, we retain only those gauge potentials satisfying $m^a A_a = 0$. Under this physically motivated restriction, both $X_a$ and $A_a$ are bona fide matter-space 1-forms satisfying $m^a X_a = 0 = m^a A_a$. Each therefore carries three independent components. Once one degree of freedom is absorbed by the matter-space gauge choice, constraints (B) and (D) are precisely sufficient to identify the two 1-forms up to a gauge transformation, so that
\begin{align} \label{FdA}
F_{ab} = \nabla_a X_b - \nabla_b X_a ~.
\end{align}
This expression automatically satisfies the matter-space constraints and represents the homogeneous sector of the electromagnetic field within the present frame.

It is physically essential to distinguish gauge fields restricted by $m^a A_a = 0$ from general unconstrained fields. 
To elucidate this distinction, consider the Aharonov-Bohm effect in the presence of a magnetic flux. 
Suppose $m^a$ points entirely along the temporal direction, establishing a frame where $\vec{E}=0$ but $\vec{B} \neq 0$.   
For a particle with charge $q$ traversing a closed spatial loop $P$, the quantum geometric phase shift is given by:  
\be{AB}
\Delta \varphi \propto  q \oint_{P} A_a dx^a .
\ee
If the integration path is deformed by $\delta x^a$, the resulting physical effect depends heavily on the direction of the variation:
\begin{itemize}
\item  A spatial variation $\delta x^i$ alters the magnetic flux enclosed by the loop, thereby modulating the phase shift.
\item  A temporal variation $\delta t$, however, cannot alter the enclosed flux because the magnetic field is strictly static in this setup. Consequently, the scalar potential $A_0$ contributes nothing to the observable geometric phase. 
\end{itemize}
This canonical example indicates that only the components of the gauge potential satisfying $m^a A_a = 0$---namely, those inherited from the matter space---contribute to the physically relevant phase. It therefore supports the view that the essential electromagnetic degrees of freedom are encoded in the matter-space fields themselves.

The analysis in this section has been purely kinematical. It identifies the geometrically preferred frame, clarifies how magnetic-type sources arise when $\bm{G}$ is not closed, and singles out the restricted gauge sector associated with the matter space. In the next section, we adopt the standard action-based viewpoint of relativistic fluid dynamics to couple this structure to ordinary electric charge and derive the remaining Maxwell equation. This does not represent a shift away from the matter-space construction of electromagnetism; rather, it shows how that construction is embedded and completed at the dynamical level within the same fluid framework.

\section{Electromagnetic dynamics from relativistic fluid dynamics} \label{sec3}

From this point onward, we restrict attention to the duality frame $F_{ab}=G_{ab}$ identified in Sec.~\ref{sec2} as the geometrically preferred realization of the matter-space construction. The analysis of Sec.~\ref{sec2} was purely kinematical: it showed how the three-dimensional matter space fixes the geometric structure of the electromagnetic field, and how the sourced equation in the dual sector can be interpreted in terms of a magnetic-type current when $\bm{dG}\neq 0$.

Ordinary electromagnetism, however, also requires the electric source sector. That sector does not follow from the kinematical matter-space constraints alone and must be supplied by additional physical input. In this section, we therefore adopt the standard action-based viewpoint of relativistic fluid dynamics in order to couple the matter-space-induced fields to conserved electric currents and derive the remaining Maxwell equation.

We stress that this variational step does not represent a shift away from the matter-space construction developed above, nor does it impose Maxwell dynamics externally. Rather, it embeds the kinematical structure obtained in Sec.~\ref{sec2} into the usual covariant fluid framework and provides its dynamical completion once electric charge carriers are included.

Accordingly, we adopt a first-order (Palatini-type) formulation in which the vector potential $A_a$ and the antisymmetric tensor $F_{ab}$ are varied as independent fields. The action is therefore not assumed \emph{a priori} to enforce their usual relation. Instead, that relation is recovered only after imposing the matter-space constraints derived in Sec.~\ref{sec2}, at which point the resulting field equation reduces to the standard sourced Maxwell equation.

We implement this construction within the covariant variational framework of relativistic fluids.
Restricting attention to the \emph{leading local} terms compatible with covariance and the (restricted) matter-space gauge symmetry, we truncate the effective action at quadratic order in the fields.
In the massless sector this excludes a Proca term $A_aA^a$, which would explicitly break gauge invariance.
Higher-derivative corrections (i.e.\ terms involving additional gradients beyond those already encoded in $F_{ab}$) are treated as subleading in an effective-field-theory sense and are therefore omitted.
Under these assumptions the unique nontrivial quadratic gauge-invariant scalar constructed from the two-form is $F^{ab}F_{ab}$.
The resulting Lagrangian should thus be viewed as the minimal dynamical completion of the matter-space kinematics once electric charge carriers are included.

The relativistic Lagrangian, a scalar in $(3+1)$-dimensional spacetime, depends on the fields introduced above and on the spacetime metric $g_{ab}$, and is given by
\begin{align} \label{lag}
\Lambda(A_a, F_{ab}, g_{ab})
= -\frac{1}{4} F^{ab} F_{ab} - j^a A_a ,
\end{align}
where $j^a$ denotes the electric current associated with the charge carriers.
At this stage it is introduced at the level of the fluid action, and its conservation will later follow from the matter-space construction of the carrier sector.
As in standard first-order formulations of electrodynamics~\cite{Verbin:2024ewl}, $A_a$ and $F_{ab}$ are treated as independent variables.
The Lagrangian itself therefore does not impose their usual relation; this relation is restored only after the constraint equations listed in Table~\ref{table1} are enforced, at which point the standard dynamical Maxwell equations emerge.

Introducing a Lagrangian displacement vector \(\xi^a\), the relationship between the Lagrangian variation \(\Delta\) and the Eulerian variation \(\delta\) is given by
\begin{equation} \label{var}
    \Delta = \delta + \mathcal{L}_{\xi},
\end{equation}
where \(\mathcal{L}_{\xi}\) denotes the Lie derivative with respect to \(\xi^a\). Here, \(\Delta\) and \(\delta\) capture changes relative to a reference configuration and variations with respect to spacetime fields, respectively. Since the Lagrangian variation of the matter space coordinates \(M^A\) vanishes, i.e., \( \Delta M^A = 0 \), we obtain:
\begin{equation} \label{eulervarMA}
     \delta M^A = - \mathcal{L}_{\xi} M^A.
\end{equation}
This relation allows the variation of the action to be expressed in terms of the displacement \(\xi^a\) rather than the flux.

The Eulerian variation of $G_{ab}$ is given by
\begin{align} \label{varFab}
\delta G_{ab} &= \delta G_{AB} (\nabla_a M^A)(\nabla_b M^B) 
\nonumber \\
&\quad  + G_{AB} \left[(\nabla_a \delta M^A)\nabla_b M^B + (\nabla_a M^A)(\nabla_b \delta M^B)\right] \nonumber \\ 
&= -\xi^c \nabla_c G_{ab} - (\nabla_a \xi^c)G_{cb} - (\nabla_b \xi^c)G_{ac}
\nonumber \\
&= -\mathcal{L}_{\xi} G_{ab}.
\end{align}
Because $F_{ab} = G_{ab}$, we get $\delta F_{ab} = - \mathcal{L}_{\xi} F_{ab}$.
Similarly, for the vector field \(A_a\), the variation takes the form
\begin{equation} \label{varAa}
\delta A_a = -\mathcal{L}_{\xi } A_a.
\end{equation}
As noted above, Eqs.~\eqref{varFab} and \eqref{varAa} are merely restatements of $\Delta F_{ab} =0$ and $\Delta A_a =0$. 

By performing a complete variation of the Lagrangian  $\Lambda(A,F)$ with respect to \textit{\textbf{both}} $F_{ab}$ and $A_a$, and utilizing the variations in Eqs.~\eqref{varFab} and \eqref{varAa}, we get
\begin{align} \label{varLag}
&\frac{\delta (\sqrt{-g} \Lambda)}{\sqrt{-g}}
= \delta \Lambda + \frac{1}{2} \Lambda g^{ab} \delta g_{ab} 
\nonumber 
 \\
&= \Pi^{ab} \left(-\mathcal{L}_{\xi} F_{ab}\right) + j^a \left(-\mathcal{L}_{\xi} A_a \right) + \left(\frac{\partial \Lambda}{\partial g_{ab}} + \frac{1}{2} \Lambda g^{ab} \right) \delta g_{ab}
\nonumber \\
&=\left[  A_e\nabla_a j^a-3\Pi^{ab}\nabla_{[e}F_{ab]} + 2F_{ea}\nabla_b \Pi^{ba} +2j^a\nabla_{[a}A_{e]} \right]\xi^e 
\nonumber \\
& \quad + \left[ \frac{\partial \Lambda}{\partial g_{ab}} + \frac{1}{2}\Lambda g^{ab} \right]\delta g_{ab} + \text{total derivatives}.
\end{align}
Here, the conjugate\footnote{We stress that $\Pi^{ab}$ is not a canonical momentum in the Hamiltonian sense, since no kinetic term for $F_{ab}$ is introduced. It plays the role of a conjugate variable in the variational formulation.} to $F_{ab}$ is 
\begin{align} \label{defofmomFab}
&\Pi^{ab} = \frac{\partial \Lambda}{\partial F_{ab}} = \left(\frac{\partial F}{\partial F_{ab}}\right)\left(\frac{\partial \Lambda}{\partial F}\right) = \frac{F^{ab}}{F} \Pi, 
\end{align}
where $\Pi \equiv \partial\Lambda/\partial F$. With respect to the Lagrangian \eqref{lag}, we get $\Pi / F =- 1/2$ and $\Pi^{ab}=- F^{ab}/2$. Finally, requiring the charge conservation associated with the external source,
\begin{align} \label{chaconserv}
    \nabla_a j^a = 0 ,
\end{align}
we obtain field equations from (\ref{varLag}), which are given by
\begin{align} \label{eqofmotion}
3 F^{ab}\nabla_{[e}F_{ab]} -2\left(\nabla_b F^{ba}\right) F_{ea} + 4  j^a \nabla_{[a} A_{e]} = 0. 
\end{align}

 Substituting the equation (\ref{FdA}) into the field equation (\ref{eqofmotion}) gives
\begin{align} \label{eqm2}
F_{ea} \left[ \nabla_b F^{ab} - j^a \right] = 0.
\end{align}
Equation~\eqref{eqm2} implies that 
\begin{equation}
F_{ea}\,\xi^a = 0,
\qquad
\xi^a := \nabla_b F^{ab} - j^a .
\label{eq:xi_def}
\end{equation}
This condition constrains $\xi^a$ to lie in the kernel of the antisymmetric
map $v^a \mapsto F^{a}{}_{b} v^b$.
A particularly simple and physically natural choice is to require\footnote{This is not the most general mathematical solution but a specific solution of physical interest. For the general cases involving the nontrivial kernel, we address at the end of this section. \label{FN6}} 
\begin{equation}
\nabla_b F^{ab} = j^a ,
\label{Maxeq2}
\end{equation}
which corresponds to setting the kernel component to zero and leads to the standard sourced Maxwell equation.
 Consequently, it becomes evident that equations \eqref{Maxeq1} and \eqref{Maxeq2} constitute Maxwell's equations describing electromagnetism, where $F_{ab}$ is interpreted as the electromagnetic field tensor and $A^a$ as the vector potential. 

The charge-conservation relation~\eqref{chaconserv}, which was required above for gauge invariance of the action, can in fact be given a natural matter-space interpretation. Introduce a 3-form field \(\bm{N} = \frac{1}{3!} N_{ABC}(N^{A}) dN^{A} \wedge dN^{B} \wedge dN^{C}\), defined on a three-dimensional charge-carrier matter space with coordinates $\{N^A\}$ (where $A = 1, 2, 3$), where the charge of a carrier is $q$.
From this 3-form one constructs the number-flow 4-vector \(n^a \equiv \frac{1}{3!} \epsilon^{abcd} N_{bcd}\), which is dual to the induced field \(N_{abc} \equiv (\nabla_a N^A)(\nabla_b N^B)(\nabla_c N^C) N_{ABC}\). 
The charge current,
\be{j:n}
j^a \equiv q n^a ,
\ee
serves as the source term in the field equations. 
Because the matter space is three-dimensional, its 3-form satisfies \(\bm{dN} = 0\), which projection to spacetime $\nabla_a n^a=0$ implies the charge-conservation equation \eqref{chaconserv} (cf.\ \cite{AnderssonNew}). 
Thus the charge conservation law emerges naturally as a direct consequence of the particle-number conservation in external source, which itself follows from the absence of 4-forms in matter space. 
This result highlights that the $U(1)$ symmetry underlying electromagnetism is intrinsically built into the matter-space formulation.

\vspace{.3cm}
More generally, however, Eq.~(\ref{eq:xi_def}) does not imply $\xi^a = 0$ when the two-form $F_{ab}$ is degenerate as noted in Footnote~\ref{FN6}, and explore the possibility of nonlinear modifications to Maxwell's equations.
In the one-fluid case considered here, the field satisfies
$\bm{G} \wedge \bm{G} = 0$ and $m^a G_{ab} = 0$ from Eqs.~\eqref{tildeff} and \eqref{orthom} because $F_{ab} = G_{ab}$,
so that the kernel of $F_{ab}$ is nontrivial.
As a result, the most general solution consistent with Eq.~(\ref{eq:xi_def})
can be written as\footnote{One might initially expect the additional term to appear as $\beta' m^a$, without the charge factor $q$. However, since the electromagnetic field tensor $F^{ab}$ changes sign under charge inversion, $q \to -q$, the additional term must likewise change sign, thereby justifying its proportionality to $q$.}
\begin{equation}
\nabla_b F^{ab} = j^a + \beta'\, m^a
= q\,(n^a + \beta m^a),
\label{Maxeq3}
\end{equation}
where \(j^a = q n^a\) as in Eq.~\eqref{j:n}, and $\beta$ is an arbitrary scalar\footnote{Notice that this non-linear contribution comes from the fact that the kernel part of $F_{ab}$ is degenerate, which is a direct consequence of the matter space construction. Therefore, this term has the same geometric origin as the $\alpha m^a$ term in Eq.~\eqref{*dG}.  } parametrizing the component of $\nabla_b F^{ab}$ along $\ker(F)$.

The choice $\beta = 0$ therefore represents a particular representative
within this kernel, selected for simplicity and to recover the conventional
form of the the standard Maxwell equations (cf.\ Eqs.~\eqref{FdA}, \eqref{Maxeq2}) that govern classical electromagnetism. 
Fixing $\beta$ in the general case requires additional physical input
beyond the constraint (\ref{eq:xi_def}), and will not be pursued further in the present work.
In contrast, Eqs.~\eqref{M^A}, \eqref{m^dphi}, and \eqref{orthom} imply that $m^a$ can be expressed nonlinearly in terms of $A^a$ and $F_{ab}$, recalling that energy densities are typically quadratic in the fields. 
Therefore, for $\beta \neq 0$, Eq.~\eqref{Maxeq3} describes a generalized class of nonlinear Maxwell equations.

We now summarize the role of the variational formulation employed in this section. The action principle introduced here does not generate the electromagnetic dynamics independently. Rather, when combined with the matter-space constraints established in Sec.~\ref{sec2}, it supplies the electric source sector and yields the familiar Maxwell equation with electric currents. In this sense, the variational formulation should be viewed not as an alternative foundation for the theory, but as the dynamical completion of the matter-space construction within the standard covariant framework of relativistic fluid dynamics.

The emergence of Maxwell's equations in this framework therefore relies crucially on the matter-space constraints. The action principle provides a convenient and covariant way to incorporate electric charge carriers and to verify that the resulting theory remains compatible with conventional electromagnetism.

At the same time, one issue remains open. In the main branch leading to Eq.~\eqref{Maxeq2}, the standard sourced Maxwell equation is recovered for the electric sector, but the status of the dual sourced term~\eqref{*dF} proportional to $\alpha m^a$ has not yet been fully resolved. In particular, when $\alpha \neq 0$ or, equivalently when $\bm{dG}\neq 0$, the relation between the Bianchi identity, magnetic-type sources, and electromagnetic duality requires further analysis. We turn to this question in the next section.

\section{Electromagnetic duality and its consequences} \label{sec4}

Let us now revisit Eq.~\eqref{*dG} and analyze the case when $\bm{ d G}\neq 0 $. This equation corresponds to a modified version of Maxwell's equations that includes a magnetic-type source, $\alpha m^a$. 

We argue that if we postulate that electromagnetism possesses the duality symmetry, $(\vec{E}, \vec{B}) \to (\vec{B}, -\vec{E})$, a particular case of the continuous dual symmetry with respect to the electric-magnetic rotation, 
\begin{equation}
\vec{E} \to \vec{E} \cos \theta + \vec{B} \sin\theta, \qquad
\vec{B} \to -\vec{E} \sin \theta + \vec{B} \cos\theta,
\end{equation}
the Bianchi identity~\eqref{Maxeq1} follows naturally from the structure of the theory, provided that no external magnetic sources (monopoles) are present.

In the presence of the $\alpha$-term, the standard Bianchi identity in Eq.~\eqref{Maxeq1} no longer holds in its conventional form. Consequently, the complementary part of Maxwell's equations is not governed by Eq.~\eqref{Maxeq2}, but rather by the modified dynamics of Eq.~\eqref{eqofmotion}.
However, this conclusion is premature.
Before reaching  this conclusion, let us apply the duality symmetry to Eq.~\eqref{Maxeq3} and investigate its implications.
To this end, consider an external magnetic source current,
\begin{equation}
j_M^a = q_M n_M^a,
\end{equation} 
where $n_M^a$ is the conserved fluid 4-vector ($\nabla_a n_M^a=0$) defined on the three-dimensional magnetic charge carrier space $N_M$, which carries magnetic charge $q_M$.
Under the duality transform $F^{ab} \to \mbox{*}F^{ab}, ~ q \to q_M, ~n^a \to n_M^a$ including the sources, Eq.~\eqref{Maxeq3} becomes   
\begin{align} \label{finalMaxwell}
\nabla_{b} \mbox{*}F^{ab} &= q_M(n_M^a + \bar \beta m^a), 
\end{align}
with $\beta \to \bar \beta$.
Comparing with Eq.~\eqref{*dG}, we get $\alpha = q_M \bar \beta$.
This relation yields an important conclusion:
even if $\bar \beta \neq 0$, in the absence of external magnetic charge carriers ($q_M=0$), the dual field has no source, i.e. $\nabla_b \mbox{*}F^{ab}=0$.
Therefore, the Bianchi identity~\eqref{Maxeq1} remains valid whenever external magnetic charge carriers are absent and the theory admits duality symmetry. 
This observation provides an alternative justification for employing Eq.~\eqref{Maxeq3}, which was originally derived under the assumption of the Bianchi identity.

It is important to emphasize that the above conclusion relies on two assumptions: (i) the theory respects duality symmetry, and (ii) no magnetic charge carriers are present. If either of these conditions fails, the argument does not hold.

When the theory lacks duality symmetry and $\alpha \neq 0$, 
the resulting Maxwell equations derived in sections~\ref{sec2} and \ref{sec3} become
\begin{align} \label{finalMaxwell}
\nabla_{b} \mbox{*}F^{ab} &= \alpha m^a, \nn \\
F_{ea} \left(\nabla_b F^{ba}  - 6 j^a_M \right) &= 2 j^a \nabla_{[a} A_{e]},
\end{align}
where we used $\nabla_{[e} F_{ab]} = 2 \epsilon_{ecab} j^c_M$. 
In contrast to the dual-symmetric case, these equations reveal an intrinsic asymmetry.
This asymmetry arises because, in our construction, electromagnetism is not a fundamental gauge field but rather \textit{emerges} dynamically from the underlying matter content of a single fluid. 
Importantly, the magnetic `current' $\alpha m^a_M$ is not an independent quantity but is directly tied to the underlying fluid motion.

When external magnetic charges are present and the theory preserves duality symmetry, Eq.~\eqref{Maxeq3} can no longer be employed.
In this situation, the fully dual symmetric formulation is required.
One natural extension is to introduce a second, dual matter space  $\mbox{*}M$, distinct from the original matter space $M$. 
In such a framework, the conventional vector potential $A^a$ would reside in $M$, while the dual potential $C^a$ would be associated with $\mbox{*}M$, allowing for a manifestly symmetric description of the  electric and magnetic sectors. 
The model Lagrangian~\eqref{lag} would then require a reformulation into a fully dual-symmetric manner~\cite{Zwanziger1971,Zwanziger1968,Kato,Bliokh}. We defer the detailed development of this extended framework for future work.

This completes the main line of argument in the frame $F_{ab}=G_{ab}$. In this formulation, the matter-space structure fixes the kinematics, the relativistic-fluid action provides the coupling to electric currents, and the role of electromagnetic duality becomes clear in determining when the standard Maxwell theory is recovered. In the physically relevant branch, the exclusion of magnetic charge restores the Bianchi identity and thereby closes the construction.

\section{Alternative  frame: $F_{ab}=-\mbox{*}G_{ab}$} \label{sec5}

For completeness and comparison, we turn to the alternative  frame 
$$
F_{ab}=-\mbox{*}G_{ab}.
$$ 
This identification is formally admissible and has the appealing feature that Eq.~\eqref{*dF} immediately takes the form of the standard sourced Maxwell equation for $F^{ab}$, with
\[
j^a \equiv \alpha m^a
\]
playing the role of an effective electric current. In this sense, the present frame gives a more direct realization of the sourced electric sector than the main construction developed in the frame $F_{ab}=G_{ab}$.

At the same time, this apparent simplicity comes at a conceptual cost. In order to recover the homogeneous Maxwell equation, one must additionally require that the dual 2-form $\mbox{*}\bm{G}$ be closed,
\[
\bm{d}\mbox{*}\bm{G}=0.
\]
Since the Hodge dual is defined only after introducing the spacetime metric and Levi-Civita tensor, this closure condition is not intrinsic to the topology of the matter space itself. Thus, unlike in the preferred frame $F_{ab}=G_{ab}$, the recovery of the Bianchi identity in the present frame does not arise directly from the matter-space geometry but instead requires an additional spacetime-level postulate.

More explicitly, once $\bm{d}\mbox{*}\bm{G}=0$ is imposed, Eq.~\eqref{Maxeq1} follows and supplies the homogeneous Maxwell equations. By Poincar\'e's lemma, the closed 2-form $F_{ab}$ can then be written locally as
\be{F:A}
F_{bc} = \nabla_b A_c - \nabla_c A_b,\nn
\ee
with $A_a$ a 1-form defined on spacetime. If, however, the condition $\bm{d}\mbox{*}\bm{G}=0$ is relaxed, then the Bianchi identity is violated and magnetic monopole configurations become possible in the usual sense.

An essential difference from the preferred frame is that the matter-space 1-form $X_a$ no longer plays a direct electromagnetic role. In the frame $F_{ab}=G_{ab}$, the matter-space structure and the restricted gauge sector together lead naturally to the identification of $X_a$ with the gauge potential up to gauge equivalence. In the present frame, by contrast, no corresponding relation emerges: the electromagnetic dynamics is expressed directly in terms of $F_{ab}$ and $A_a$, while $X_a$ remains conceptually detached from the resulting field equations. For this reason, the present frame is less tightly connected to the intrinsic matter-space geometry, even though it remains formally consistent.

\section{Summary and discussion} \label{sec6}

We have reformulated electromagnetism within the matter-space framework commonly used in relativistic fluid dynamics. Our central premise is that the essential electromagnetic structure should be encoded in matter-space fields and their spacetime pull-backs. From this viewpoint, we argued that the physically relevant gauge sector is the restricted one satisfying $m^aA_a=0$, and that the Aharonov--Bohm phase is naturally associated with this matter-space gauge potential.

Starting from a three-dimensional Euclidean matter space equipped with 0-, 1-, and 2-form fields, we derived the orthogonality relations in Eqs.~\eqref{M^A}, \eqref{m^dphi}, and \eqref{orthom}, together with the constraint equations summarized in Table~\ref{table1}. Among the two admissible duality frames, we identified the frame $F_{ab}=G_{ab}$ as the preferred one, because it preserves the most direct geometric relation between the physical field strength and the intrinsic matter-space 2-form. In this frame, the matter-space constraints determine the kinematical structure, while the action-based formulation of relativistic fluid dynamics provides the coupling to electric charge carriers and yields the remaining Maxwell equation.

A central point of the present construction is that the Bianchi identity is not simply assumed from the outset, but must be understood through the internal structure of the theory. In the preferred frame, its status is controlled by the closure properties of $\bm{G}$ and, ultimately, by consistency with electromagnetic duality in the absence of magnetic charge. By contrast, the alternative frame $F_{ab}=-\mbox{*}G_{ab}$ remains formally admissible but is less intrinsic from the standpoint of matter-space geometry, because recovering the Bianchi identity requires the additional spacetime-level condition $\bm{d}\mbox{*}\bm{G}=0$, which is not  dictated by the topology of the matter space.
\begin{table}[htp]
\caption{Retrospective comparison of the two duality frames.}
\label{table2}
\small
\renewcommand{\arraystretch}{1.80}
\setlength{\tabcolsep}{4pt}
\begin{tabular}{|c|c|c|}
\hline
\parbox[c]{0.22\linewidth}{\centering \textbf{Feature}} &
\parbox[c]{0.34\linewidth}{\centering \textbf{$F_{ab}=G_{ab}$}} &
\parbox[c]{0.34\linewidth}{\centering \textbf{$F_{ab}=-\mbox{*}G_{ab}$}} \\
\hline
\parbox[c]{0.22\linewidth}{\centering Sourced equation} &
\parbox[c]{0.34\linewidth}{\centering From fluid action} &
\parbox[c]{0.34\linewidth}{\centering From matter-space constraint} \\
\hline
\parbox[c]{0.22\linewidth}{\centering Bianchi identity} &
\parbox[c]{0.34\linewidth}{\centering From closure / duality} &
\parbox[c]{0.34\linewidth}{\centering From $\bm{d}\mbox{*}\bm{G}=0$} \\
\hline
\parbox[c]{0.22\linewidth}{\centering Role of $X_a$} &
\parbox[c]{0.34\linewidth}{\centering Gauge potential up to gauge} &
\parbox[c]{0.34\linewidth}{\centering No direct EM role} \\
\hline
\parbox[c]{0.22\linewidth}{\centering From $\nabla_a m^a=0$} &
\parbox[c]{0.34\linewidth}{\centering Helicity conservation} &
\parbox[c]{0.34\linewidth}{\centering Charge conservation} \\
\hline
\parbox[c]{0.22\linewidth}{\centering Aharonov--Bohm phase} &
\parbox[c]{0.34\linewidth}{\centering From restricted matter-space $A_a$} &
\parbox[c]{0.34\linewidth}{\centering No analogous interpretation} \\
\hline
\parbox[c]{0.22\linewidth}{\centering Electric source} &
\parbox[c]{0.34\linewidth}{\centering Charge carriers in fluid action} &
\parbox[c]{0.34\linewidth}{\centering Effective source from matter space} \\
\hline
\parbox[c]{0.22\linewidth}{\centering Magnetic source} &
\parbox[c]{0.34\linewidth}{\centering From non-closure of $\bm{G}$} &
\parbox[c]{0.34\linewidth}{\centering From violation of $\bm{d}\mbox{*}\bm{G}=0$} \\
\hline
\parbox[c]{0.22\linewidth}{\centering One-fluid restriction } &
\parbox[c]{0.34\linewidth}{\centering degeneracy of electromagnetic field } &
\parbox[c]{0.34\linewidth}{\centering absent} \\
\hline
\end{tabular}
\end{table}
The resulting picture is therefore twofold. On the one hand, the frame $F_{ab}=G_{ab}$ provides the geometrically preferred realization of electromagnetism emerging from matter-space structure. On the other hand, the frame $F_{ab}=-\mbox{*}G_{ab}$ remains a useful alternative formulation, whose comparison with the preferred frame helps clarify which aspects of the theory are intrinsic to matter space and which enter only after spacetime dualization.
The comparison between the two duality frames are summarized  in Table~\ref{table2}.

\vspace{.3cm}
We now turn to several interpretive aspects of the theory, including the role of the vector field $m^a$, its relation to helicity, and possible extensions beyond the minimal one-fluid construction.

We first clarify the role of the vector field $m^a$ in the one-fluid realization of electromagnetism. Given a field configuration $F_{ab}$ and a gauge potential $A_a$ (up to gauge scalars $\{\phi\}$), the direction of $m^a$ is fixed by the orthogonality conditions, such as $m^aF_{ab}=0$, while its overall normalization is determined, up to a constant factor, by the conservation law $\nabla_a m^a=0$. Thus $m^a$ should be regarded not as an independent field, but as a derived quantity determined by $F_{ab}$ and $A_a$.

In particular, the orthogonality condition
\begin{equation}\label{KelHelA}
m^aF_{ab}=0 
\quad\Longrightarrow\quad
m^a\nabla_{[a}A_{b]}=0,
\end{equation}
is formally analogous to the force-free condition $v^a\omega_{ab}=0$ in ideal fluid dynamics, where $v^a$ denotes the fluid 4-velocity and $\omega_{ab}=2\nabla_{[a}v_{b]}$ its vorticity. By analogy, one may regard $A_a$ as a flow potential along $m^a$, with $F_{ab}$ playing the role of the associated vorticity. Equation~\eqref{KelHelA} then implies that this vorticity is Lie-transported, or ``frozen in,'' along the flow generated by $m^a$, in close analogy with the Kelvin--Helmholtz theorem. In this sense, Eq.~\eqref{orthom} encodes the conservation of electromagnetic vorticity along the matter flow. Equivalently, when the flow is twist-free, i.e. when $\nabla_{[a}m_{b]}=0$, the gauge potential is Lie-conserved:
\begin{equation}\label{LiewrtmofA}
\pounds_{m}A_b = m^a\nabla_a A_b + A_a\nabla_b m^a = 2\nabla_{[a}m_{b]}\,A^a =0.
\end{equation}
Hence the line integral $\oint A_a\,dx^a$ around loops comoving with $m^a$ is conserved. More generally, even when \(\nabla_{[a}m_{b]}\neq0\), the combination of \(\pounds_{m}A_b\) with the condition \(m^aA_a=0\) implies that the circulation of \(A_a\) around comoving loops remains conserved up to a gradient contribution.

A natural question is which electromagnetic quantity plays the role of fluid vorticity in the present framework. 
Combining the two equations (B) and (D) in Table~\ref{table1} and defining the helicity 4-vector,
\begin{align} \label{defJ}
    \mathcal{H}^a \equiv \epsilon^{abcd} A_{[b} F_{cd]} = A_b \mbox{*}F^{ba}~,
\end{align}
we derive another nontrivial conservation equation:
\begin{align} \label{defcurrent}
    \nabla_a \mathcal{H}^a = 0.
\end{align}
Although this expression appears gauge dependent, helicity becomes gauge invariant and physically meaningful for closed or periodic systems, or under suitable boundary conditions such as $\vec B\cdot\hat n=0$ on the boundary, where $\hat n$ denotes the outward unit normal.
Notice that explicit calculation gives 
\be{dH}
 \nabla_a \mathcal{H}^a = F_{ab} \mbox{*}F^{ab} = - 4 \vec{E} \cdot \vec{B}.
 \ee
 This result shows that helicity conservation is a direct consequence of the one-fluid structure encoded in Eq.~\eqref{tildeff}.
  The vector $\mathcal{H}^a$ satisfies
\begin{align} \label{orthoH}
\mathcal{H}^a \nabla_a \phi = 0, 
\quad 
\mathcal{H}^a A_a = 0,
\quad 
\mathcal{H}^a F_{ab} = 0.
\end{align}
Consequently, up to an overall normalization constant, it is natural to identify $m^a \propto \mathcal{H}^a$, so that the helicity current itself provides the vorticity vector in the electromagnetic context\footnote{Dimensional remark: since $\nabla_a\mathcal{H}^a$ has units of energy density (as $F_{ab}{} \mbox{*}F^{ab}$ does), $\mathcal{H}^a$ carries dimensions of action per unit volume (or angular momentum density), consistent with its interpretation as helicity flux.}.

In this framework, the flow vector $m^a$ is expressed directly in terms of the electric and magnetic fields, $\vec{E}$ and $\vec{B}$. We perform a standard $3+1$ decomposition of the field strength tensor as follows:
\begin{equation} \label{eq:3+1decomp}
F_{0i} = E_i,\quad F_{ij} = \varepsilon_{ijk} B^k\,.
\end{equation}
From this, the components of $m^a$ are derived as
\begin{align} \label{eq:mwrtEB1}
m^0 &\propto \mathcal{H}^0 = \epsilon^{0ijk} A_i F_{jk} = A_i B^i\,,
 \nn  \\
m^i &\propto \mathcal{H}^i = \epsilon^{i0jk} A_0 F_{jk} + \epsilon^{ij0k} A_j F_{0k} \nn \\
     &  \quad \quad ~
     	= A_0 \epsilon^{ijk} B_k + \epsilon^{ijk} A_j E_k\,.
\end{align}
In the Coulomb gauge ($A_0 = 0$), the expressions simplify to
\begin{align} \label{eq:mwrtEB2}
m^0 \propto \vec{A} \cdot \vec{B}\,, \qquad
\vec{m} \propto \vec{A} \times \vec{E} \,.
\end{align}
Here, $\mathcal{H}^0 = \vec{A} \cdot \vec{B}$ corresponds to the magnetic helicity density, while $\vec{\mathcal{H}} = \vec{A} \times \vec{E}$ represents the helicity flux. Consequently, Eq.~\eqref{defcurrent} reduces to the continuity equation
\begin{align} \label{eq:continuity}
\frac{\partial}{\partial t}(\vec{A}\!\cdot\!\vec{B})
+\nabla\!\cdot(\vec{A}\times\vec{E})
=0\,,
\end{align}
indicating that $\mathcal{H}^a$ is conserved and gauge-invariant under appropriate boundary conditions. Thus, identifying $m^a$ with a suitably scaled $\mathcal{H}^a$ naturally interprets the fluid flow vector as a \emph{helicity current} within the one-fluid model.
Not all electromagnetic fields can be realized in this one-fluid framework; more general configurations require multi-fluid descriptions or different matter-space structures.  In future work, one could explore relaxations of these assumptions to classify broader classes of field configurations.

We now briefly comment on a possible nonlinear extension of the theory and its implications for the self-energy problem.
 This nonlinear correction in Eq.~\eqref{Maxeq3} is analogous to mass renormalization effects, reminiscent of the motivation behind Born-Infeld electrodynamics~\cite{BIEM}.
Because Eq.~\eqref{Maxeq3} contains nonlinear contributions, it is interesting to examine the electron self-energy problem. 
When $\beta = 0$, the electron self energy problem appears as follows:
The electric energy density of a static charge $q$ located at $r=0$ is 
\begin{align} \label{energydenb=0}
u(r) \;=\; \frac{\epsilon_0 E_r^2}{2} \;=\; \frac{q^2}{32\pi^2 \epsilon_0\,r^4},
\end{align}
where $\epsilon_0$ is the electric permittivity of free space. So integrating over volume, the total electric energy becomes
\begin{align} \label{totenergyb=0}
U(\delta) = \int_{r>\delta} u\,dV \sim \frac{q^2}{8\pi\epsilon_0\,\delta},
\end{align}
where $\delta$ denotes the characteristic size of the electron. Taking the infinitesimal limit $\delta \to 0$, the energy diverges. To compensate for this divergence, one should choose the bare mass of electron to have negative infinite value. 

The divergence of the self-energy disappears when $\beta  >0$. 
In this case, the solution to Eq.~\eqref{Maxeq3} is 
\begin{align} \label{Coulombfieldbneq0}
E_r = \frac{q}{4\pi \epsilon_0 r^2(1+ \frac{ r_c}{ r})}, \qquad
 r_c \equiv \frac{\beta q^2}{8\pi}
\end{align}
with all other components vanish. Note that when $r \gg r_c $, the formula reproduces the $\beta=0$ result.
The electric potential is modified around $r \sim r_c$.  
In high energy $e^+e^-$ scattering experiments, the Coulomb potential is kept to $10^{18} $m~\cite{Christy:2021snt,Neyenhuis,Tu}. 
This result presents an upper bound on the value of $\beta$:
$$
r_c = \frac{\beta \rm{e}^2}{8\pi} < 10^{-18}\rm{m}. 
$$
Near \(r\to0\), however, the denominator modifies the divergence so that the energy density 
\begin{align} \label{energydenbneq0}
u(r) = \frac{\epsilon_0 E_r^2}{2} =\frac{q^2}{32\pi^2 \epsilon_0 r^4(1+ \frac{\beta q^2}{8\pi r})^2} 
\end{align}
over any volume be finite. Integrating over the whole space, the total energy stored in the electro-static field is 
\begin{align} \label{totenergybneq0}
U_{\rm total} = \int_0^\infty 4\pi r^2 u(r)\,dr = \frac{1}{\beta \epsilon_0} = \frac{q^2}{8\pi \epsilon_0 r_c }. 
\end{align}
If we take this value to be the electron mass, we have $r_c \sim 1.4 \,{\rm fm}$, which is comparable to the nuclear size and slightly smaller than the so-called classical electron radius, $2.8\, {\rm fm}$~\cite{Haken2005}. 
Note that $r_c$ does not denote the size of electron because the electric charge in this model is located at the center $r=0$, which makes the electric field diverge there.
Therefore, this result is consistent with the known upper bound of the electron size $10^{-22}\,{\rm m}$.
Even though this result is better than that of the classical electrodynamics, it fails to explain the self-energy problem in the classical level because it does not satisfy the previous bound.
Therefore, this theory still requires the quantum mechanical treatment to resolve the self-energy problem.

Beyond the classical level, a natural next step is to extend the matter-space viewpoint to quantum regime. 
Although the present work is formulated at the classical level, the matter-space construction suggests a possible route toward quantizing the electromagnetic field directly in terms of matter-space degrees of freedom. 

In the present work, we have focused on identifying the minimal geometric and kinematical ingredients required for the emergence of electromagnetic structure from relativistic fluid dynamics. The constraints summarized in Table~\ref{table1} constitute a complete set implied by the dimensionality and gauge structure of the matter space. Rather than implementing these constraints through Lagrange multipliers in the action, we have imposed them at the level of the equations of motion. This choice allows us to keep the formulation minimal and to emphasize the geometric origin of the Maxwell equations, without introducing the additional technical machinery of a full constrained (Dirac-type) analysis.

A systematic implementation of these constraints via Lagrange multipliers, together with a Hamiltonian or phase-space formulation, would provide a more comprehensive description of the theory and is left for future work. In this broader context, it will also be important to explore how spinor fields and additional symmetries can be incorporated within the matter-space framework.

\section*{Acknowledgements}
This work was supported by the National Research Foundation of Korea(NRF) grant with grant number RS-2023-00208047 (H.K.,J.H.,Y.Y.), RS-2019-NR040081 (J.H.), RS-2025-00553127 (J.H.), and NRF-2020R1F1A1068410 (J.L.).


\begin{thebibliography}{}

\bibitem{Brandenburg} A. Brandenburg and E. Ntormousi, "Galactic Dynamos", Annu. Rev. Astron. Astrophys. 61 (2023) 561, arXiv: 2211.03476[astro-ph.GA], \hreff{https://doi.org/10.1146/annurev-astro-071221-052807}

\bibitem{Subramanian} K. Subramanian, ``The origin, evolution and signatures of primordial magnetic fields'', Rep. Prog. Phys. 79 (2016) 076901, arXiv:1504.02311[astro-ph.CO], \hreff{https://doi.org/10.1088/0034-4885/79/7/076901}

\bibitem{Davidson} P. A. Davidson, ``An Introduction to Magnetohydrodynamics'', Cambridge University Press (2001), \hreff{https://doi.org/10.1017/CBO9780511626333} 

\bibitem{Kulsrud} R. M. Kulsrud and E. G. Zweibel, 
``On the Origin of Cosmic Magnetic Fields",  Rept. Prog. Phys. 71 (2008) 0046091, arXiv:0707.2783[astro-ph], \hreff{https://doi.org/10.1088/0034-4885/71/4/046901}

\bibitem{Pariev} V. I. Pariev and S. A. Colgate, ``A Magnetic Alpha-Omega Dynamo in Active Galactic Nuclei Disks: I. The Hydrodynamics of Star-Disk Collisions and Keplerian Flow", Astrophys. J. 658 (2007) 114, arXiv:astro-ph/0611139, \hreff{https://doi.org/10.1086/510734}; ``A Magnetic Alpha-Omega Dynamo in Active Galactic Nuclei Disks: II. Magnetic Field Generation, Theories and Simulations", Astrophys. J. 658 (2007) 129, arXiv:astro-ph/0611188, \hreff{https://doi.org/10.1086/510735}

\bibitem{Hernandez} J. Hernandez and P. Kovtun, ``Relativistic magnetohydrodynamics", JHEP 1705 (2017) 001, arXiv:1073.0875[hep-th], \hreff{https://doi.org/10.1007/JHEP05\%282017\%29001}

\bibitem{Green} M. B. Green, J. H. Schwarz and E. Witten, ``Superstring Theory'', Cambridge University Press (2012), \hreff{https://doi.org/10.1017/CBO9781139248563}

\bibitem{Rovelli} C. Rovelli, ``Quantum Gravity'', Cambridge University Press (2004), \hreff{https://doi.org/10.1017/CBO9780511755804}

\bibitem{Jacobson} T. Jacobson, ``Thermodynamics of Spacetime: The Einstein Equation of State", Phys. Rev. Lett. 75 (1995) 1260, arXiv:gr-qc/9504004, \hreff{https://doi.org/10.1103/PhysRevLett.75.1260}

\bibitem{Padmanabhan} T. Padmanabhan, ``Emergent Gravity Paradigm: Recent Progress", Mod. Phys. Lett. A 30 (2015) 03n04, 1540007, arXiv:1410.6285[gr-qc], \hreff{https://doi.org/10.1142/S0217732315400076}

\bibitem{Lee} J. Lee and H. S. Yang, ``Quantized K$\ddot{a}$hler Geometry and Quantum Gravity", J. Korean Phys. Soc. 72, (2018) 1421, \hreff{https://doi.org/10.3938/jkps.72.1421}; J. Lee and H. S. Yang, " Dark energy and dark matter in emergent gravity",  J. Korean Phys. Soc. 81 (2022) 910, arXiv:1709.04914 [hep-th], \hreff{https://doi.org/10.1007/s40042-022-00605-9}

\bibitem{Taub54}
A. H. Taub, ``General Relativistic Variational Principle for Perfect Fluids", Phys. Rev. {\bf 94} (1954) 1468, 
\hreff{https://doi.org/10.1103/PhysRev.94.1468}

\bibitem{Carter72} 
B. Carter and H. Quintana, ``Foundations of General Relativistic High-Pressure Elasticity Theory", Proceedings of the Royal Society of London. Series A, Mathematical and Physical Sciences {\bf 331} (1972) 57, \hreff{https://doi.org/10.1098/rspa.1972.0164}

\bibitem{Carter73}
B. Carter, ``Elastic perturbation theory in General Relativity and a variation principle for a rotating solid star'', Commun. Math. Phys. {\bf 30} (1973) 261. 
\hreff{https://doi.org/10.1007/BF01645505}

\bibitem{Carter89}
B. Carter, ``Covariant theory of conductivity in ideal fluid or solid media'', Lect. Notes Math. 1385 (1989) 1, \hreff{https://doi.org/10.1007/BFb0084028}

\bibitem{AnderssonNew}
N. Andersson and G. L. Comer, ``Relativistic fluid dynamics: physics for many different scales'', Living Rev. Rel. 10 (2007) 1, arXiv:gr-qc/0605010,
 \hreff{https://doi.org/10.12942/lrr-2007-1}

\bibitem{LK2022}
H.~C.~Kim and Y.~Lee, ``Heat conduction in general relativity'', Class. Quantum Grav. 39 (2022) 245011, arXiv:2206.09555 [gr-qc], \hreff{https://doi.org/10.1088/1361-6382/aca1a1}

\bibitem{Kim:2023lta}
H.~C.~Kim, ``Steady heat conduction in general relativity'', PTEP \textbf{2023} (2023) 5, 053A02, arXiv:2302.03291 [gr-qc], \hreff{https://doi.org/10.1093/ptep/ptad062}


\bibitem{Dubovsky:2011sj}
S.~Dubovsky, L.~Hui, A.~Nicolis and D.~T.~Son,
Phys. Rev. D \textbf{85}, 085029 (2012)
[arXiv:1107.0731 [hep-th]].
\hreff{https://doi.org/10.1103/PhysRevD.85.085029}

\bibitem{Verbin:2024ewl}
Y.~Verbin, B.~Pulice, A.~{\"O}vg{\"u}n and H.~Huang,
Phys. Rev. D \textbf{111}, no.8, 084061 (2025)
doi:10.1103/PhysRevD.111.084061
[arXiv:2412.20989 [gr-qc]].

\bibitem{Zwanziger1971}
D. Zwanziger, "Local Lagrangian quantum field theory of electric and magnetic charges", Phys. Rev. D {\bf 3}, 880 (1971), arXiv:hep-th/0106277, \hreff{https://doi.org/10.1103/PhysRevD.3.880}

\bibitem{Zwanziger1968}
D. Zwanziger, "Exactly Soluble Nonrelativistic Model of Particles with Both
Electric and Magnetic Charges", Phys. Rev. {\bf 176}, 1480 (1968), \hreff{https://10.1103/PhysRev.176.1480}; "Quantum field theory of particles with both electric and magnetic charges", Phys. Rev. {\bf 176}, 1489 (1968), \hreff{https://doi.org/10.1103/PhysRev.176.1489}

\bibitem{Kato} A. Kato and D. Singleton, "Gauging Dual Symmetry",  Int.J.Theor.Phys. {\bf 41}, 1563 (2002), arXiv:hep-th/0106277, \hreff{https://doi.org/10.1023/A:1020192616580}

\bibitem{Bliokh} K.~Y. Bliokh, A.~Y. Bekshaev, and F. Nori, "Dual electromagnetism: Helicity, spin, momentum, and angular momentum", New J.Phys. {\bf 15}, 033026 (2013), arXiv:1208.4523 [physics.optics], \hreff{https://doi.org/10.1088/1367-2630/15/3/033026}, corrigendum: New J.Phys. {\bf 18}, 089053 (2016)

\bibitem{BIEM}
M. Born and L. Infeld, (1934). "Foundations of the New Field Theory", Proceedings of the Royal Society of Edinburgh Section A, 144 (1934) 425, \hreff{https://doi.org/10.1098/rspa.1934.0059}

\bibitem{Christy:2021snt}
M.~E.~Christy, T.~Gautam, L.~Ou, B.~Schmookler, Y.~Wang, D.~Adikaram, Z.~Ahmed, H.~Albataineh, S.~F.~Ali and B.~Aljawrneh, \textit{et al.}, ``Form Factors and Two-Photon Exchange in High-Energy Elastic Electron-Proton Scattering,'' Phys. Rev. Lett. \textbf{128}, 102002 (2022), arXiv:2103.01842 [nucl-ex], \hreff{https://doi:10.1103/PhysRevLett.128.102002}.

\bibitem{Neyenhuis}
B.~Neyenhuis, D.~Christensen, and D.~S.~Durfee, "Testing nonclassical theories of electromagnetism with ion interferometry."
Phys. Rev. Lett. \textbf{99}, 200401 (2007), arXiv:phys/0606262, \hreff{https://doi:10.1103/PhysRevLett.99.200401}.

\bibitem{Tu}
L.-C.~Tu and J.~Luo, ``Experimental tests of Coulomb's Law and the photon rest mass,'' Metrologia {\bf 41} (2004) S136, 
 \hreff{https://doi:10.1088/0026-1394/41/5/S04}


\bibitem{Haken2005}
Haken, H., Wolf, H.C. and Brewer, W.D. (2005) The Physics of Atoms and Quanta. ``Advanced Texts in Physics," Springer, Berlin, Heidelberg, p. 69,
\hreff{https://doi.org/10.1007/3-540-29281-0} 


\end{thebibliography}
\end{document}